# 衬底上石墨烯制备及改性研究*

田 圆 赵倩莹 胡 靖 周 辰 缪 灵** 江建军
(华中科技大学电子科学与技术系 武汉 430074)

**摘 要** 大面积高质量石墨烯的制备对石墨烯电子特性及石墨烯基纳器件相关研究有重要意义。本文综述了近几年来衬底上制备石墨烯的相关实验以及衬底与石墨烯相互作用研究的重要进展。目前,采用化学气相沉积、外延生长等方法可在衬底表面上制备出较大面积、高质量的石墨烯材料。衬底与石墨烯相互作用和界面间晶格匹配、原子成键及电荷转移等密切相关,其对吸附石墨烯的几何结构、能带结构及电子特性等产生明显影响。实验与理论计算的结合可望加深衬底与石墨烯作用机理的理解,指导衬底上石墨烯制备及改性的进一步研究。

# Preparation and Modification of Graphene on Substrate

*Tian Yuan Zhao Qianying Hu Jing Zhou Chen Miao Ling* Jiang Jianjun*
(Department of Electronic Science and Technology HuaZhong University of Science and Technology, Wuhan 430074, China)

**Abstract** Since the successful isolation of the single atomic layer of graphite in 2004, graphene has drawn great interests due to its unique properties, including high mechanical strength, outstanding conductivity, high coefficient of thermal conductivity, etc. It is significant to manufacture large-scale and high-quality graphene on various substrates for the study of the characteristics of graphene and the research of the nano-devices basing on graphene. This paper selectively reviews recent experiment advances in graphene made on different substrates, SiC, $SiO_2$, Cu, Ni, Co, Ru, for instance. Nowadays, we can obtain large area of high quality graphene by using different methods, such as CVD, epitaxial growth, mechanical separation, etc. We can manufacture graphene on nonmetals including SiC, GaAs, $SiO_2$, and metals covering Cu, Ni, Co, Ru, Au, Ag, etc. This article especially reviews the interaction between the graphene and the substrates. The mechanism of interaction is closely related to the mismatch of the lattice, weakness of the bonds, the transformation of the electrons between the few layer graphene and substrates. Also, the interaction between them has great influences on geometry, energy band, and coefficient of thermal conductivity, phonon dispersion, optical waveguide performances and the properties of electrons of the graphene. The combination of the experiment and the calculation (such as density functional theory, tight-binding method, molecular dynamics simulation, etc.) can make a deeper understanding of the mechanism of the effects between graphene and different substrates, which can be served as a guide for further study.

收稿: 2011 年 月, 收修改稿: 2011 年 9 月
* 国家自然科学基金 (No. 50771047)资助。
** Corresponding author e-mail: miaoling@hust.edu.cn

## Contents

## 1 引言

2004 年发现石墨烯后[1]，由于其独一无二的优良物理性质，迅速成为了物理学、化学和材料学等领域的研究热点[2,3]，其中最引人注意的便是它在纳电子器件中的应用前景。石墨烯为六角晶格整齐排布的二维周期蜂窝状结构，它是构成其他石墨材料的基本单元，可以翘曲成富勒烯，卷成碳纳米管或堆垛成石墨[2]。石墨烯是当前世界上已发现的最薄、最坚硬、最有韧性的物质（其杨氏模量可以达到 1 000 GPa 以上[4]），具有很高的结晶度和稳定性[5]，热导率可接近甚至超过钻石[6]。石墨烯具有特殊的能带结构，其价带 $\pi$ 电子和导带 $\pi^*$ 电子相交于费米能级[7]。能带交叠点附近，电子由于受到周围对称晶格势场的影响，有效质量变为零，类似于相对论的狄拉克电子。因此，这些交叠的非等效点通常被称为狄拉克点。

石墨烯独特的电子能带结构导致其拥有许多新奇的电学性质，其电子迁移率 $\mu$ 可达15 000 $cm^2 \cdot V^{-1} \cdot s^{-1}$[8]，远远超过了电子在金属导体或半导体中的移动速度。而且只与杂质和缺陷有关，而与温度无关，这意味着 $\mu$ 还可以大幅度提高。可利用石墨烯的高电子迁移率用于弹道输运晶体管（ballistic transistor）的制作，并制备出晶体管原型器件[9]。石墨烯双极场效应明显，通过栅电压调制，载流子可以在电子和空穴之间连续性的过渡，载流子浓度 n 可达到 $10^{13}$ $cm^{-2}$。当石墨烯通过化学修饰[10—12]、物理吸附[13—15]或切割成石墨烯纳米带[16—22]时，可调控其电子结构及带隙以拓宽其应用领域。

基于石墨烯诸多良好性质，石墨烯材料在很多领域中有极大应用前景[23]，高质量、大尺寸的石墨烯材料的可控制备，对开展石墨烯本征物理特性及其应用研究，都具有重要的科学与技术上的价值[3]。目前，制作石墨烯的方法主要有：微机械剥离法[24]，晶体外延生长法[25,26]，过渡金属催化的化学气相沉积法[27—37]，氧化还原法（Standenmaier 法[38]，Brodie 法[39]，Hummers 法[40]）等。相对而言，微机械剥离法可通过较为简单的方式制备高质量的石墨烯，但是比较费时费力，其较大的精度控制难度和较差的重复性导致这种方法只适用在实验室制备石墨烯。化学气相沉积法以金属晶体作为衬底，可通过衬底的选择、生长的温度、前驱物的暴露量等生长参数对石墨烯的生长进行较好的调控[41]。晶体外延生长法和过渡金属催化的化学气相沉积法有比较好的制备效果，可获得较大面积、高质量的石墨烯，但是石墨烯的生长机理仍未完全探明。

石墨烯与衬底的相互作用对其制备生长及各种性质影响非常明显，非常有必要对其深入研究。这不仅对探求更好的制备高质量、大面积石墨烯的方法，还是有选择的优化改性石墨烯性质以适用于各领域，都显得格外重要。本文将着重从以下三方面来介绍相关领域的研究进展：(1)在衬底表面上制备石墨烯的最新实验研究；(2)石墨烯与不同衬底的作用机理；(3)衬底对石墨烯的改性作用的研究。

## 2 衬底表面制备石墨烯的实验研究

可用做石墨烯衬底的材料种类很多，粗略分类为非金属类衬底（包括 SiC、$SiO_2$、GaAs 等）和金属类衬底（包括 Cu、Ni、Co、Ru、Au、Ag 等）。实际上，石墨烯衬底的分类并没有明确的界限。目前已经从单一种类物质做衬底过渡到由多种物质在结构上或者性质上相互配合形成掺杂衬底或多层衬底，且获得了良好的效果。但由于非金属类与金属类的衬底表面和石墨烯的作用机理差别明显，因此为便于研究将衬底分为以上两类。

### 2.1 石墨烯在非金属衬底上生长

由于 SiC 和 $SiO_2$ 衬底材料可以利用微电子领

域标准化的刻蚀工艺(nanolithography)加工,所以人们首先探究了石墨烯在 SiC 和 $SiO_2$ 衬底上的生长。比较有代表性的是 de Heer 小组[42,43]通过氧化或者氢刻蚀来预处理 6H-SiC(000$\bar{1}$)单晶表面,在超高真空下加热至1 000 ℃去除表面氧化物,再加热样品 SiC,从而在表面外延生长石墨烯。然而采用这种外延生长制备石墨烯的方法所得到的石墨烯以岛状生长为主,石墨烯层数波动较大、缺陷很多,难以获得大面积厚度均一的石墨烯片层。和机械分离法比较起来,外延生长的方法制备的石墨烯有较高的载流子迁移率特性,但是观测不到量子霍尔效应。Emtsev 等[44]则利用常压在 SiC(0001)表面的 Si 终端生长石墨烯。这种方法获得的石墨烯弥补了先前用超高真空的方法获得的石墨烯面积过小的缺陷,同时获得的石墨烯在 $T = 27$K 下的电子迁移率可达到2 000$cm^2 \cdot V^{-1} \cdot s^{-1}$。2010 年,De Heer 组又利用逐级光刻和微纳工艺在 SiC 衬底的($1\bar{1}0n$)面自组织(self-organized)生长窄到 40 nm 的石墨烯纳米带[45],如图 1。通过这种工艺制作出的石墨烯原型器件在低温(4 K)下具有量子限域效应,从而石墨烯能带能够产生带隙且不对石墨烯的其他电子特性产生影响,同时这种石墨烯器件中载流子迁移率在室温下可达到2 700 $cm^2 \cdot V^{-1} \cdot s^{-1}$。而且,他们在 0.24 $cm^2$ 的 SiC 片上集成了10 000个石墨烯晶体管,是目前报道最大集成密度。

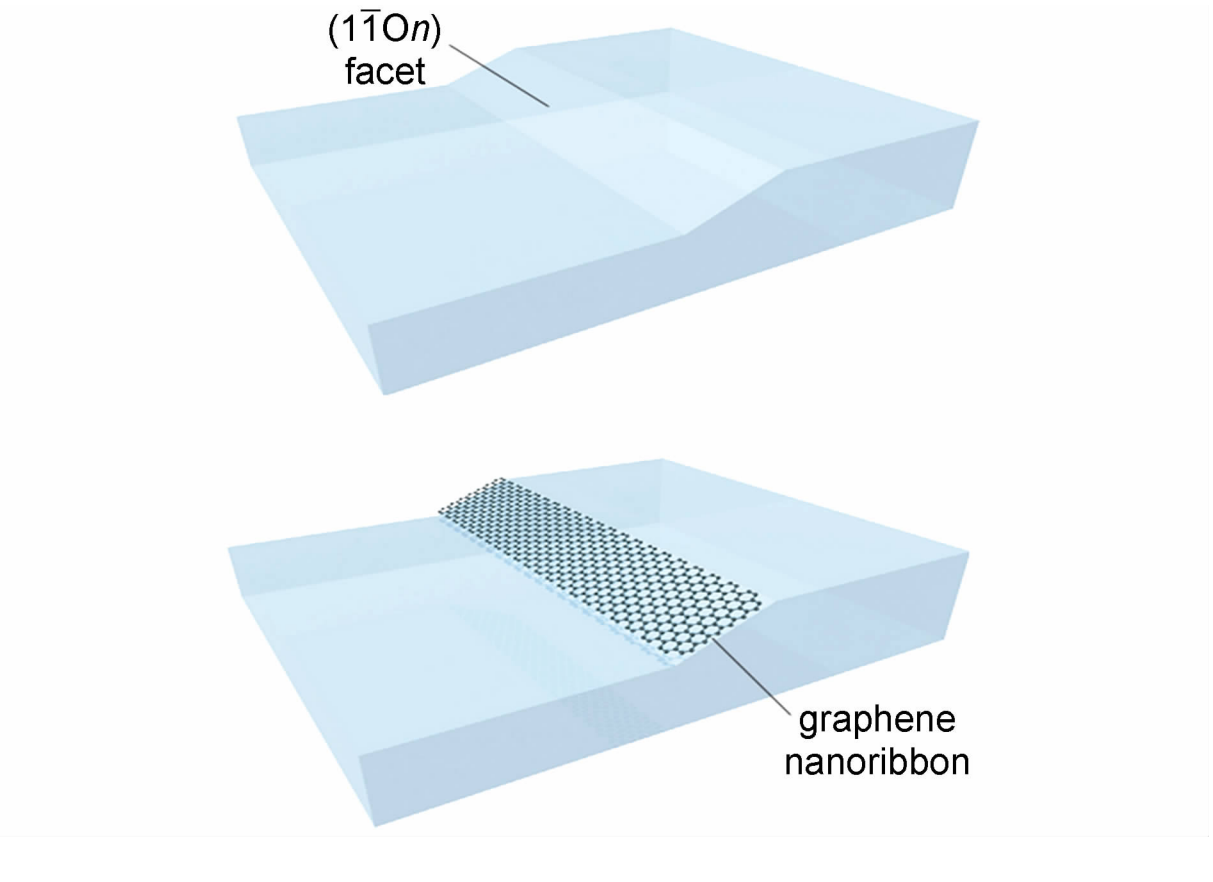

**图 1**　SiC 晶体表面台阶处生长石墨烯示意图[45]

**Fig. 1**　Process for tailoring of the SiC crystal for selective graphene growth and device fabrication[45]

近年研究结果表明,$SiO_2$ 同样可与其他金属材料相配合生长出完整的石墨烯[46]。纵观石墨烯在不同非金属结构上生长情况可知,对于 SiC 结构,石墨烯可生长的表面有 Si 表面和 C 表面;对于 $SiO_2$ 来说,石墨烯可生长的表面有 Si 表面和 O 表面。由于 SiC 和 $SiO_2$ 的晶格结构完全不同,两种物质的 Si 表面与石墨烯的作用也不完全一样。

### 2.2　石墨烯在金属衬底上生长

金属衬底上制备的石墨烯具有高质量、高致密度以及面积大的特点。在金属衬底上制备石墨烯已经成为目前生长石墨烯的主要方法之一。

2008 年,Yu 等[47]用表面偏析法在金属 Ni 衬底上制备了高质量的石墨烯,并可以控制石墨烯薄膜的厚度和缺陷。随着这种方法的发展,单层石墨烯或者是双层石墨烯已经可以覆盖 87% 的 Ni 的表面[48]。同年,Coraux 等[35]利用低压气相沉积法在 Ir(111)表面生长了单层石墨烯,扫描隧道显微镜的研究结果证实此石墨烯结构可以跨越金属台阶,在 Ir(111)表面上形成连续、低缺陷密度、微米尺度的单层碳结构。

2009 年,Ruoff 小组[49]在 Cu 薄膜表面上成功地制备了达到厘米级的大面积、高质量石墨烯,而且所获得的石墨烯主要为单层结构。他们接着针对任意衬底研究了石墨烯薄膜转移过程,发现室温下,在硅/二氧化硅衬底上制作的双门控场效应晶体管的电子迁移率可以高达4 050 $cm^2 \cdot V^{-1} \cdot s^{-1}$。高鸿钧等[50]在Ru(0001)表面上通过表面偏析的方法获得毫米量级的单晶石墨烯。Kong 小组[48]和 Kim 等[51]分别实现了在多晶 Ni 薄膜上外延生长厘米量级石墨烯。Juang 等[52]利用 SiC 作为 C 源,同时采用电子束蒸发将 200 nm 厚的 Ni 置于 SiC 上。最后对整个体系快速升温至 750 ℃后迅速冷却到室温,在这个过程中,大量的 C 原子偏析到 Ni 上面,在冷却的过程中,石墨烯比较容易的从 Ni 表层剥离。

2010 年,Fonin 等[53]在 Rh(111)表面生长石墨烯。根据 Rh(111)表面的单层石墨烯 STM 图像显示,扩展域单晶石墨生长没有任何大规模缺陷。2011 年,Sun 等[54]在金属Ir(111)衬底上生长石墨烯,虽然存在由于晶格不匹配导致的 moire 皱褶(35 ± 10 pm),石墨烯片层仍可以在 Ir 衬底表面大片且高质量地生长。

综上所述,无论是关于非金属衬底的研究还是关于金属衬底的研究,近年来都取得了很大的进展。研究人员在追求更高质量、更大面积的目标中不断努力。

## 3　不同衬底与石墨烯的作用机理

生长在衬底上的石墨烯与衬底表面直接接触,其相互作用强弱决定了石墨烯的不同性质。研究衬

底表面与石墨烯的不同作用机理,可以更深入的研究石墨烯-衬底体系的性质。对于非金属衬底,主要是界面间原子成键作用;对于金属衬底,主要是晶格匹配和电子转移的作用;对台阶状的衬底主要是台阶附近的衬底原子的活性;对多层石墨烯-衬底结构,则主要是最接近衬底的石墨烯层的缓冲作用。

### 3.1　石墨烯-非金属衬底作用机理

在非金属衬底上制备出的石墨烯呈现出很好的电子特性和载流子高迁移率,为将石墨烯应用于纳电子器件提供了可能途径。到目前为止,已经有关于 SiC、$SiO_2$、Si、GaAs[55]、BN[56, 57]等多种非金属材料被尝试用作石墨烯的生长衬底,其与石墨烯的相互作用主要由界面之间的成键情况决定。一般而言,未钝化的衬底表面原子较为活泼,石墨烯会与之形成共价键,对于 H 钝化后的衬底表面,石墨烯较难与之成键;而这些共价键会对石墨烯的几何结构和电子特性产生显著影响。

#### 3.1.1　石墨烯-SiC 作用

SiC 和石墨烯的作用主要分为 Si 表面和 C 表面两种类型。Kim 等[58]建立了 $6\sqrt{6}\times6\sqrt{3}R30°$的表面原子结构模型,并通过第一性原理计算方法研究了 SiC 衬底的 Si 表面与单层石墨烯以及双层石墨烯的相互作用。对于单层石墨烯,C 原子和衬底 Si 原子之间形成了强共价键,单层石墨烯的能带图上清晰的出现了 σ 键能带,而费米能级附近的 π 键能带消失了。Hiebel 等[59]将 6H-SiC(000$\bar{1}$)表面在真空环境中热处理,发现由于 C 表面原子悬挂键的强作用,使石墨烯表面空间结构发生了变化,形成了 2 × 2、3 × 3 等不同的构型。Seubert 等[60]对 SiC 表面构型进行了研究,提出了 2 × 2 构型的原子结构模型。Magaud 等[61]根据 Seubert 等提出的 SiC 2 × 2 原子结构模型,通过密度泛函理论计算了 2 × 2 构型的 SiC 表面与石墨烯的相互作用。C 表面由于内层 Si 原子的作用发生变化而被钝化,与石墨烯作用减弱。吸附其上的石墨烯性质类似于非支撑石墨烯,狄拉克点仍在费米能级。

#### 3.1.2　石墨烯-$SiO_2$ 作用

和 SiC 类似,$SiO_2$ 衬底存在 Si 和 O 两种不同的表面,其中 Si 表面的化学性质并不活泼。Kang 等[62]通过第一性原理计算发现,石墨烯与未钝化 Si 表面之间距离较远,作用非常微弱,衬底并没有对石墨烯的电子结构造成明显影响,与用 H 原子完全钝化掉表面 Si 原子两个悬挂键后的情况非常类似。而在钝化表面 Si 原子单个悬键后,石墨烯中的 C 原子和衬底 Si 原子形成了很强的Si—C共价键,表明衬底表面的悬键活性对二者之间的成键作用起着至关重要的作用。

而 $SiO_2$ 的 O 表面的化学性质则比较活泼。Shemella 等[63]研究了 $SiO_2$ 的 O 表面与单层、双层石墨烯的相互作用,以及 H 钝化氧原子后的表面与单层石墨烯之间的作用。石墨烯片层与衬底 O 表面形成较强共价键,其能带结构与非支撑石墨烯的半金属性能带差别很大。当 $SiO_2$ 的 O 表面上生长双层石墨烯时,石墨烯的能隙减小。当 O 表面被 H 钝化后形成羟基组,石墨烯层与衬底之间没有成键,如图 2,因此为了在 $SiO_2$ 衬底上生长类似非支撑能带结构的石墨烯,$SiO_2$ 的氧表面须被 H 钝化。

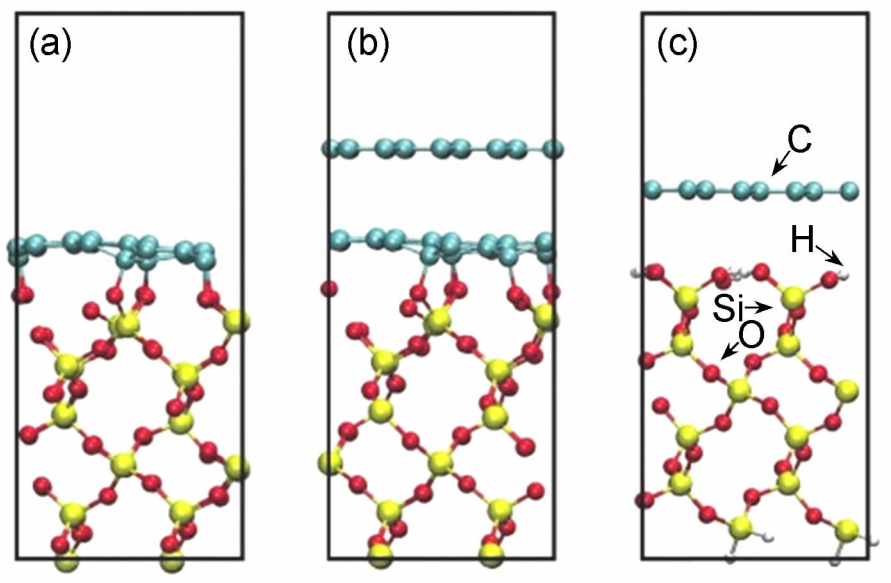

**图 2**　石墨烯在 $SiO_2$ 衬底 O 表面的结构模型图[63]

**Fig. 2**　Graphene sheets above the $SiO_2$ substrate with O-termination and without H-passivation for (a) the monolayer and (b) the bilayer. The H-passivated system (c) is shown and no covalent bonding between graphene and substrate was observed[63]

#### 3.1.3　石墨烯-Si 衬底作用

石墨烯-Si 衬底作用规律与石墨烯-$SiO_2$ 的情况类似。对于没有 H 钝化的 Si 表面而言,由于表面原子活性比较大,与石墨烯会形成 σ 共价键,对石墨烯的电子轨道排布产生影响。当 Si 表面被 H 钝化后,石墨烯的电子特性不会发生显著的改变[64]。

### 3.2　石墨烯-金属衬底作用

石墨烯的金属衬底的种类较多,作用规律各不相同。可大致将石墨烯-金属作用分为以下三类:(1)衬底表面较惰性,不会与石墨烯发生反应而改变石墨烯的特性;(2)衬底能与石墨烯形成弱的相互作用,但并没有成键,在较大间距时可不改变石墨烯原本的能带特性;(3)衬底表面的化学性质比较活泼,易与石墨烯发生反应,需要隔离或钝化措施来保持石墨烯的良好的电子特性。另外,金属衬底和石墨烯之间的吸附与两者间的晶格匹配程度紧密相关,匹配度越高,石墨烯的吸附也就相对更强,两者

之间也更易成键。

3.2.1　晶格匹配对石墨烯-金属衬底作用的影响

Wang 等[57] 利用密度泛函理论计算了在 Ru(0001)面外延生长的石墨烯的特性。Ru 表面生长的石墨烯有褶皱起伏不平，造成褶皱的重要原因就是石墨烯层的晶格常数和衬底金属的晶格常数有微小的不同。同时石墨烯和衬底之间会由于褶皱而形成不同的区域。在强作用区域，石墨烯能带中出现能隙。他们同时研究了 Ir 衬底上生长的石墨烯，并指出 Ir-石墨烯之间的作用力仅为 Ru-石墨烯的三分之一，解释了在 Ir 上生长的石墨烯褶皱结构不甚明显的原因。

3.2.2　惰性金属衬底

金属 Au 衬底的表面不活泼，在其上生长石墨烯不会改变石墨烯的能带特性。Varykhalov 等[65] 探讨了不同的重金属衬底表面上掺杂石墨烯能带的改变。对于 Cu 和 Ag 而言，石墨烯能带会打开，形成禁带。对于 Au 而言，即便生长的石墨烯重掺杂，石墨烯的能带也不会打开，石墨烯仍然保持着良好的半金属导电性能。除此之外，Giovannetti 等[66] 利用密度泛函理论计算金属衬底对石墨烯的掺杂改性，发现 Al、Ag 和 Pt 对石墨烯的影响较弱，石墨烯独特的能带结构得以保留，费米面的上下平移在 0.5 eV 以内，其数值可用金属的功函数来解释，图 3。

3.2.3　微作用金属衬底

Cu 金属衬底与石墨烯的反应程度居中。Xu 和 Buehler 等[67] 用第一性原理研究了 Cu(111) 和 Ni(111)分别和单层石墨烯接触的界面。通过计算石墨烯和 Cu(111) 表面的结合能，top fcc ( 面心立方)结构是最稳定的结构。当 Cu 原子和 C 原子距离比较大时，费米能级在石墨烯的 π 和 π * 能带之间，如图 4。

3.2.4　强作用金属衬底

Ni、Ir 等强作用衬底和石墨烯的晶格匹配程度比较高，石墨烯容易在这些衬底上生长。Xu 和 Buehler 等[67] 指出，由于开放的 d 层轨道强耦合作用，Ni 表面会与石墨烯形成较强的作用，其结合能远大于 Cu 表面和石墨烯的结合能。Brako 等[68] 通过范德瓦尔斯密度泛函方法的计算研究了石墨烯生长在Ir(111)结构上的性质，由于石墨烯和 Ir 之间晶格并非完全匹配，所以它们之间成键的键能不大。但当石墨烯的上方再加一层 Ir 形成三明治结构时，Ir 和石墨烯之间形成了较强的 $sp^3$ 杂化键。强的相互作用会破坏石墨烯的平面几何结构。Lahiri 等[69]

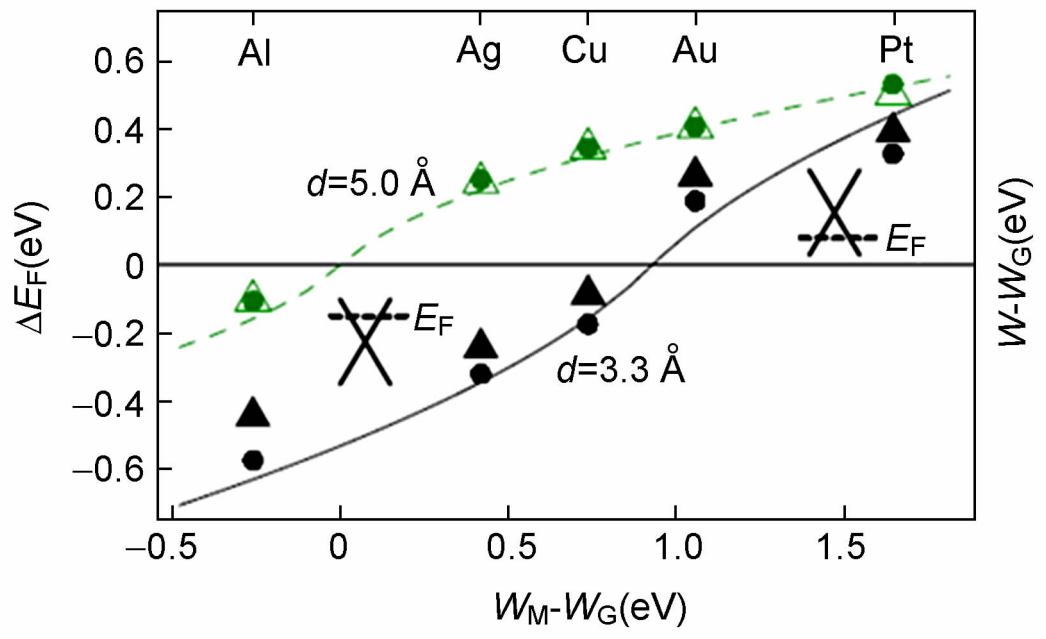

**图 3**　与石墨烯作用时不同金属衬底的功函数图[66]

**Fig. 3**　Calculated Fermi energy shift with respect to the conical point, $\triangle E_F$ (dots), and change in the work function $W$-$W_G$ (triangles) as a function of $W_M$-$W_G$, the difference between the clean metal and graphene work functions. The lower (black) and the upper [gray (green)] results are for the equilibrium ( ~3.3 Å) and a larger (5.0 Å) separation of graphene and the metal surfaces, respectively. The solid line and the dashed line follow from the model of Eq. (1) with $\triangle c = 0$ for $d = 5.0$ Å. The insets illustrate the position of the Fermi level with respect to the conical point[66]

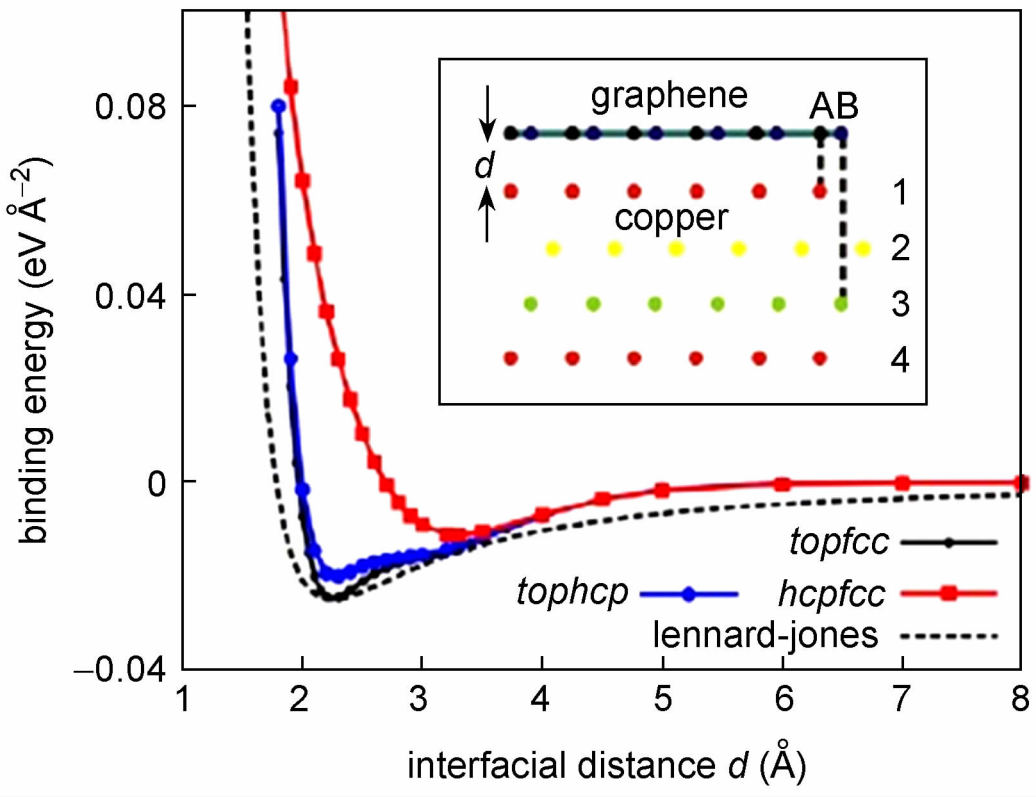

**图 4**　石墨烯和 Cu(111)表面的结合能随距离关系[67]

**Fig. 4**　Binding energy of the interface between a grapheme monolayer and copper (111) surface as a function of interface distance d. Inset: atomic structure of the interface and metal substrate represented by four atomic layers. There are three configurations with different stacking orders following the terminology, where the two carbon atoms (A and B) in the graphene unit cell cover metal atoms in layers 1 and 3 (topfcc), 1 and 2 (tophcp) or 2 and 3 (hcpfcc). The topfcc configuration has the lowest energy and is thus the most stable one[67]

探讨了 Ni-石墨烯-Ni 三明治结构，密度泛函理论计算表明，其结构不稳定。由于 Ni 和 Ni 之间的作用力太强，石墨烯会在两层金属的上方继续生长，中间

石墨烯结构被破坏。

### 3.3　金属台阶衬底的作用

采用 CVD 方法时,石墨烯可在金属表面台阶上较好生长。由于石墨烯边缘与过渡金属衬底表面的相互作用较强[70,71],石墨烯会在金属表面生长过程中形成岛状圆形区域[72]。McCarty 等[73]揭示了在低碳浓度的情况下,石墨烯的形成倾向于发生在金属台阶边缘附近;而在高碳浓度下,则倾向于形成在台阶边缘和平台上。Coraux 等[28]利用低压化学气相沉积方法在Ir(111)上面生长了单层的高纯度石墨烯,在衬底表面台阶处,石墨烯依然能够地毯式的越过台阶生长。

早在 2002 年,Bengaard 等[74]就指出在衬底表面台阶处石墨片层和衬底原子的结合力比较大,这或许就是石墨片层能够自发生长的重要原因。Saadi 等[71]利用密度泛函理论计算研究了不同过渡金属台阶状表面的石墨烯成核机理,建立了石墨烯的生长模型。在与石墨烯晶格匹配较好的金属台阶衬底上,石墨烯能够较好生长且成核大小较小。Ding 等[75]分别计算了在波纹状和台阶状金属表面石墨烯的边界形成能、石墨烯的形成能以及石墨烯成核速度,指出石墨烯在金属台阶状衬底表面成核要优于波纹状表面。由于台阶附近原子活性比较大,石墨烯在台阶边界的形成能比较小,C 团簇更容易在台阶附近聚团进而形成较大石墨烯结构,如图 5。

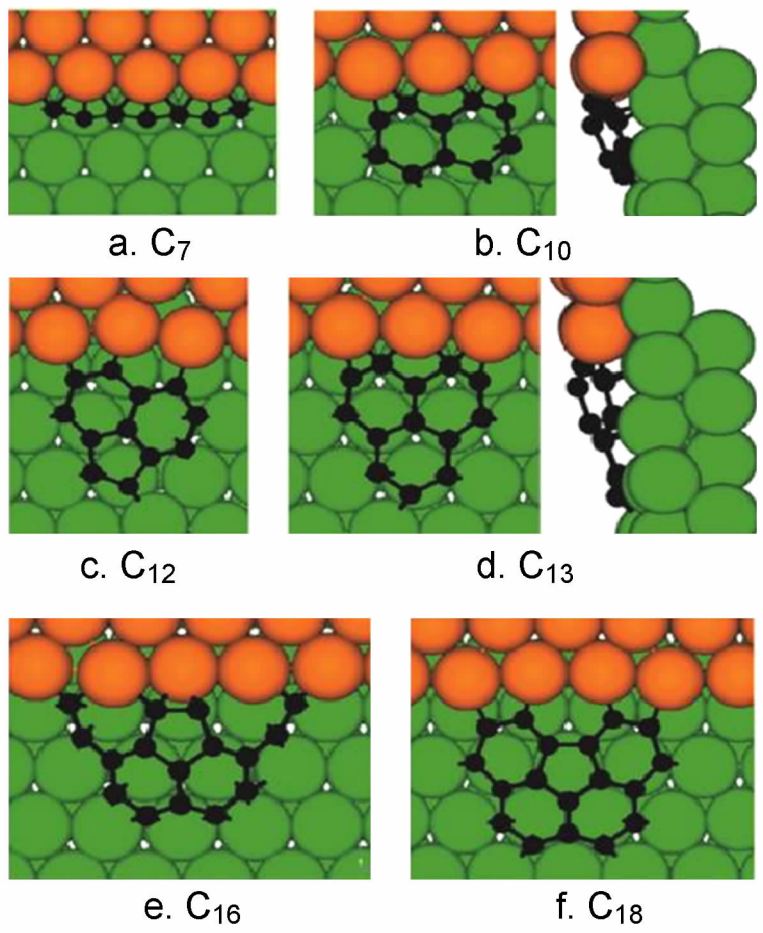

**图 5**　C 团簇在 Ni 台阶状表面聚集的结构图[75]

**Fig. 5**　Typical optimized structures of supported $C_N$ clusters near a Ni step edge on the (111) surface. Both the top and side views of the $C_{10}$ and $C_{13}$ clusters are shown[75]

目前有关石墨烯在衬底上成核机理的实验文献比较丰富[37,70—73,76—78],理论研究相对较少[71,75,79]。研究石墨烯的成核机理对指导制备更高质量更大规模的石墨烯有重要意义,相信这会是今后的研究热点。

### 3.4　多层石墨烯和衬底之间的作用

对衬底上多层石墨烯电子特性的研究指出,一般而言,对于较活泼的衬底表面,界面处第一层石墨烯扮演着缓冲层的角色,基本失去了单层石墨烯的电子特性,而第二层具有类似非支撑石墨烯的特性。当石墨烯形成三层 Bernal 堆积结构后,石墨烯能隙就会在狄拉克点处打开[80](如图 6)。De Heer 小组研究发现[81],在 SiC 和多层石墨烯之间存在一层中介层,他们命名为第 0 层(layer-0),正是由于第 0 层的存在,导致外层石墨烯出现褶皱。另有研究表明,SiC 的取向和第一层缓冲层的作用有着密切的联系[61]。实验方面,Hong 等[82]已经制备出 FLG-FET,并测出了其特性曲线,与理论研究相符。

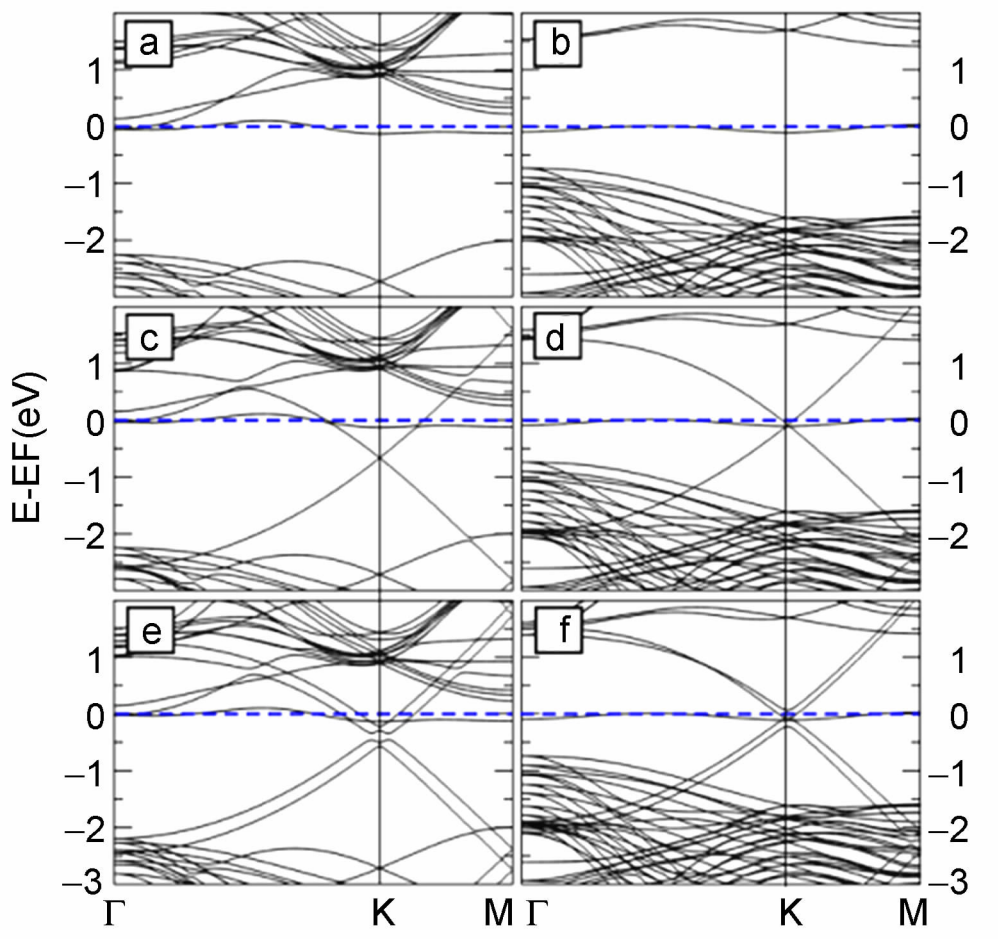

**图 6**　SiC 衬底表面多层石墨烯能带结构图[80]

**Fig. 6**　(color on line) dispersion curves for one (a,b), two (c,d) and three (e,f) C layers on bulk truncated SiC. a, c, e correspond to Si terminated face; b, d, f to C terminated face. Inset in c shows a zoom of the anticrossing in the vicinity of $E_F$[80]

## 4　衬底对石墨烯的改性作用

非支撑石墨烯拥有很多优良电子特性,如零质量的载流子、圆锥形的 π 价带和 π * 导带、狄拉克锥结构、高载流子迁移率[83]以及石墨烯纳米带能带调制作用[84]等,都给下一代纳电子器件发展提供了良好平台。由于衬底作用的引入,衬底上石墨烯的几何结构发生了较大改变,结构的改变影响电子输运特性,进而对石墨烯基纳电子器件性能有很大影响。

### 4.1　衬底对石墨烯结构的影响

石墨烯的电子特性与其特殊二维结构密切相关,每个石墨烯六元环有三个 σ 共价键及离域性较强的 π 键。衬底对石墨烯几何结构的影响主要体现在以下 3 个方面:一是衬底表面与石墨烯成键的影响;二是石墨烯生长的边界形态的影响;三是晶格匹配度对金属-石墨烯结构的影响。

Ishigami 等[85]利用 STM 观察在非金属 $SiO_2$ 衬底上生长的石墨烯。他们发现,由于在制备过程中丙烯酸光阻剂的存在,会对石墨烯结构产生不可控的扰动。一旦去除了这些杂质,生长石墨烯的表面比较光滑。Jia 等[86]发现了石墨烯纳米带边界的有效结构重建过程,而 Girit 等[87]则利用亚埃分辨率透射电镜研究了石墨烯边界生长,边缘可呈现 Z 字形(zigzag)和扶手椅型(armchair)结构,并给出了一种精确可控制备石墨烯纳米带方法。Sutter 等[88]利用低能电子显微镜观察了生长在 Pt 上的石墨烯结构,Pt 台阶上的石墨烯可以平滑生长而生长在 Ru 上的石墨烯则会具有褶皱[89—91]。这是因为石墨烯和 Ru 衬底的晶格匹配度远高于 Pt 衬底,较高的晶格匹配度意味着衬底和石墨烯之间的作用力比较大并导致褶皱的出现。

## 4.2　衬底对石墨烯带隙的影响

较强的衬底作用会使石墨烯的能带打开带隙。2007 年 Lanzara 等首先指出 SiC 基底与石墨烯间的相互作用将打开石墨烯电子结构的带隙[92]。Varchon 等利用第一性原理方法研究了 SiC 基底与石墨烯的相互作用机理[80]。Giovannetti 等[56]利用第一性原理计算方法研究了和石墨烯晶格匹配的 BN 衬底对石墨烯能带的影响,发现 BN 衬底使石墨烯打开的 53 meV 的能隙比 Cu(111)面生长的石墨烯打开的能隙大。对于 $SiO_2$ 衬底上石墨烯纳米带的研究表明,纳米带边缘与衬底表面成键方式决定了其能带和带隙大小,如图 7 所示。[93]

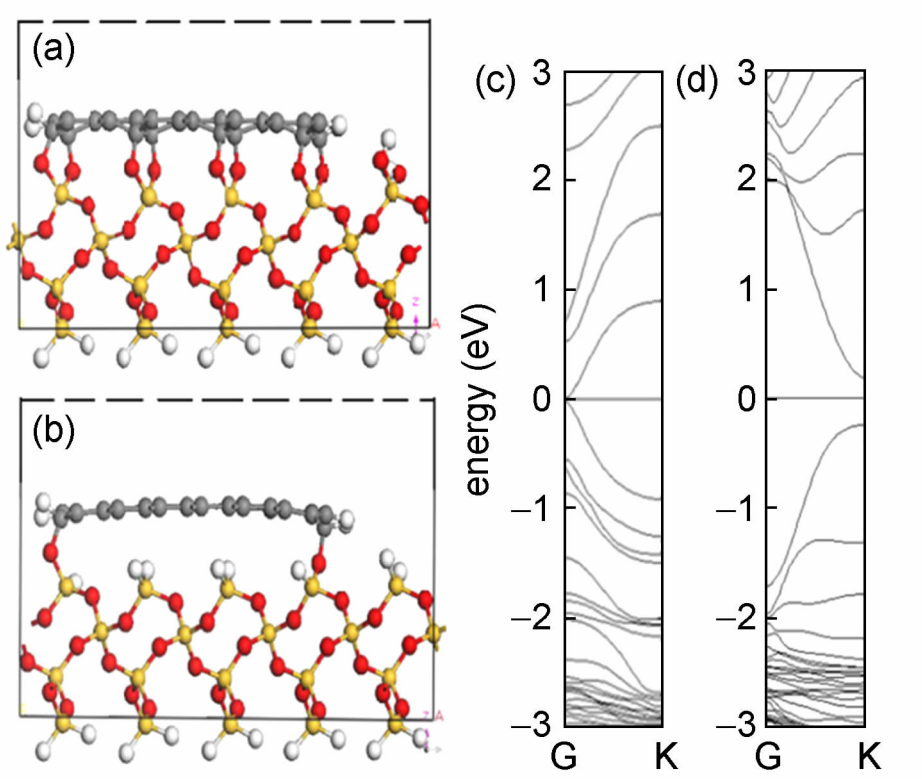

**图 7**　$SiO_2$ 衬底上石墨烯纳米带结构模型和电子能带图[93]

**Fig. 7**　(a) Structure for 7-H-ZGNRs on O1 surface. The red, yellow, gray, and white balls denote the O, Si, C, H atoms, respectively. Band structure of (b) 7-H-ZGNRs on O1 surface, (c) 7-H-ZGNRs on O2 surface and (d) 8-H-ZGNRs on O2 surface. Fermi levels are set to zero[93]

对于金属衬底而言,Giovannetti 等[66]的研究表明,不同的金属衬底对石墨烯能带有不同的影响,其中对于 Al、Ag、Cu 等金属衬底,石墨烯的能带呈现出 n 型半导体的特征,而对于 Au、Pt 等金属衬底,石墨烯的能带呈现出 p 型半导体的特征。即在未来微纳器件制备当中,可以通过金属衬底作用直接调节石墨烯电子特性。

对于金属衬底而言,不匹配的晶格会使得石墨烯产生能隙,但这种能隙主要是由于晶格的取向造成的。McCarty 等[76]发现,不同的石墨烯晶格取向在 Pt(111)上没有显著影响石墨烯的电子特性。但在石墨烯和金属衬底作用较大的体系中,如石墨烯-Pd 体系,衬底晶格取向显著地影响了石墨烯电子特性。

生长在Ni(111)、Ru(0001)以及Ir(111)衬底表面的石墨烯电子结构已有较多研究。在这些体系里,石墨烯和衬底的晶格矢量在同一个平面的同一条直线上,称为 R0 取向[94]。对于有强作用的石墨烯-衬底体系,比如 Ni 和 Ru 衬底,π 能级会降低,导致在 K 点和 K' 点上能带会打开一个间隙[42,95];而对于弱作用的石墨烯-衬底体系,比如 Ir 衬底[36],石墨烯和衬底之间的作用可以忽略,生长在这种衬底的石墨烯的电子特性和单层无衬底的石墨烯的电子特性类似。但是由于石墨烯和 Ir 之间的晶格不匹配,导致石墨烯独特的狄拉克圆锥状能带打开一条微小的带隙。经过深入的研究,Starodub 等[94]指出,在费米能级附近,R0 石墨烯的 π 能级和金属 Ir 的 5 d 能级产生较强的杂化作用,这种作用导致了石墨烯的 π 和 π * 能级的分离。接着,Elean 等探究了将 R0 石墨烯晶格方向旋转 30°之后成为 R30 石墨烯的电子特性。与 R0 石墨烯相比,R30 石墨烯在费米能级附近和金属 Ir 能带之间的杂化作用变弱,能带没有打开且石墨烯为 *p* 掺杂[95—99]。

## 4.3　衬底-石墨烯体系的热接触

衬底作用对石墨烯的导热性能也有显著影响[100],界面和杂质会降低石墨烯的热导率。考虑到集成电路应用方面,对金属、Si、$SiO_2$ 衬底研究较多。

Seol 等[101]研究表明，尽管石墨烯与 $SiO_2$ 之间存在声子的界面散射作用，外延生长在 $SiO_2$ 上的单层石墨烯的室温热导率 $k$ 仍约为 600 $W \cdot m^{-1} \cdot K^{-1}$，比其他常用薄膜电子材料的热导率高很多。他们得到 $SiO_2$ 衬底上单层石墨烯的电导率 $\sigma$ 和热导率 $k$ 与温度的关系如图 8 所示。

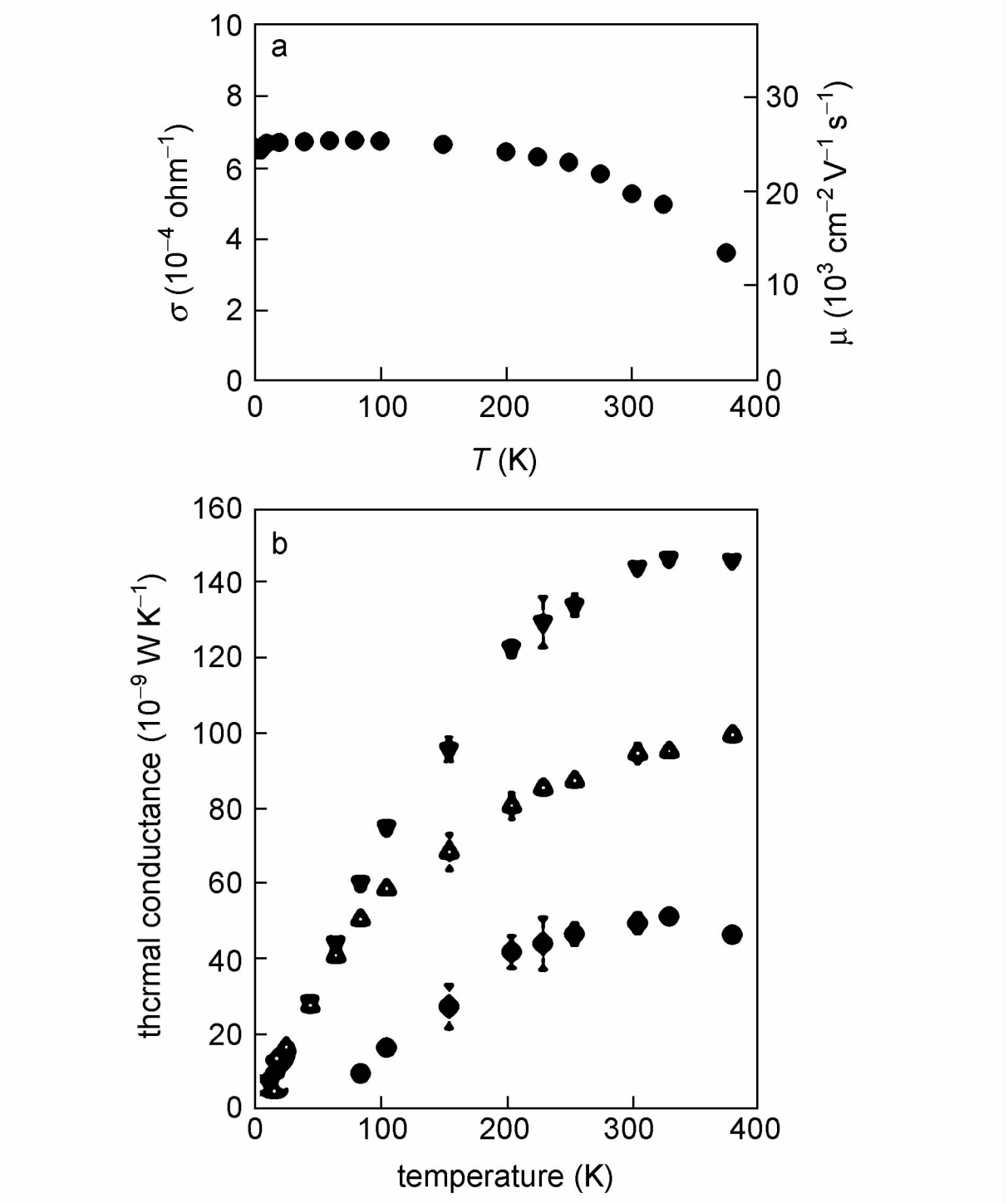

**图 8**　$SiO_2$ 衬底上单层石墨烯的(a)电导率和(b)热导率与温度的关系[101]

**Fig. 8**　(a) Measured two-probe electrical conductivity (s) and extracted electron mobility (m) of G2 as a function of temperature. (b) Measured thermal conductance of G2 before (solid downward triangles) and after (unfilled upward triangles) the SLG was etched, with the difference being the contribution from the SLG (circles)[101]

Gao 等[102]通过实验的方法，分别测量 G(石墨烯)/Ni(111)、G/Ru(0001)和G/Pt(111)的热电势。他们发现G/Pt(111)具有正的热电势，而对于 G/Ru(0001)，当探头接触样品表面时热电势为正，而在探头压在表面以下时，热电势为负。对于 G/Ni(111)则存在负的热电势，还给出纯 Pt 衬底的热电势也是负值。

### 4.4　衬底对石墨烯的其他影响

衬底对石墨烯其他特性还有许多影响，比如磁电特性、声子特性、光学特性等。这些可以指导今后的石墨烯改性研究，进一步扩大了石墨烯纳器件的应用范围。

#### 4.4.1　衬底上石墨烯带的磁电效应

Zhang 等[103]对衬底上的石墨烯纳米带的磁电特性作了细致的研究，他们预测了石墨烯纳米带在硅衬底上的磁电效应。第一性原理计算结果表明，偏压能够产生强烈的线性磁电效应，驱使载流子在纳米带和衬底间转移，改变磁场边缘处的交换分离。并且，由偏压导致的纳米带从 $n$ 型到 $p$ 型的转变能改变体系的磁电效应系数，使之由负转正。这为在非金属磁性系统的磁电效应耦合打开了一条新路。

Varykhalov 等[104]研究了生长在 Au 改性后 Ni 衬底上的石墨烯的电磁特性，生长在 Ni 上面具有准非支撑型的石墨烯可以分离出不同自旋方向的电子，具有比较高的自旋电子分离效率。如图 9。

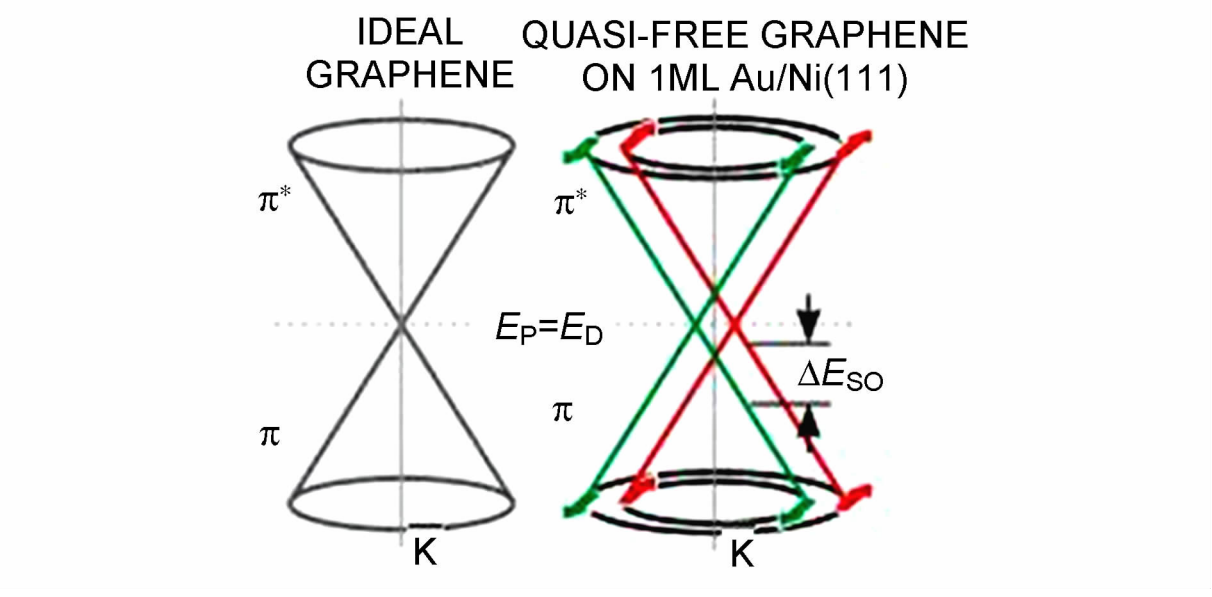

**图 9**　Ni 衬底表面具有准非支撑型的石墨烯可以分离出不同自旋方向的电子[104]

**Fig. 9**　Model for the spin-polarized fermi surface of graphene[104]

#### 4.4.2　衬底-石墨烯声子特性

石墨烯上电子与声子的耦合具有重要作用，比如影响拉曼散射和电子输运等。石墨烯与衬底间的相互作用可以改变这种耦合作用(如图 10)。在 Ni(111)面上生长的石墨烯，完全抑制了科恩反常[105]。由于在石墨烯的 π 键和 Ni 的 d 轨道之间出现了强烈的杂化，最高光学分支在 Γ 和 K 的附近完全变平。对于其他金属衬底，石墨烯同金属衬底间的距离越大，杂化越弱，科恩反常被轻微削弱。所以通过声子散射实验，可以判断石墨烯同不同金属衬底间相互作用的强弱。

#### 4.4.3　衬底-石墨烯体系的光学性能

石墨烯和不同衬底材料组成的体系具有不同的光学性能。Wlasny 等[106]利用菲涅尔光学理论计算得出，利用 7 nm 厚 Au 层衬底或者是利用 110 nm 厚 $SiO_2$ 衬底(如图 11)，石墨烯具有较好的光学性能，其光学对比度较高(可达 60%)，可以用作制作光波导。

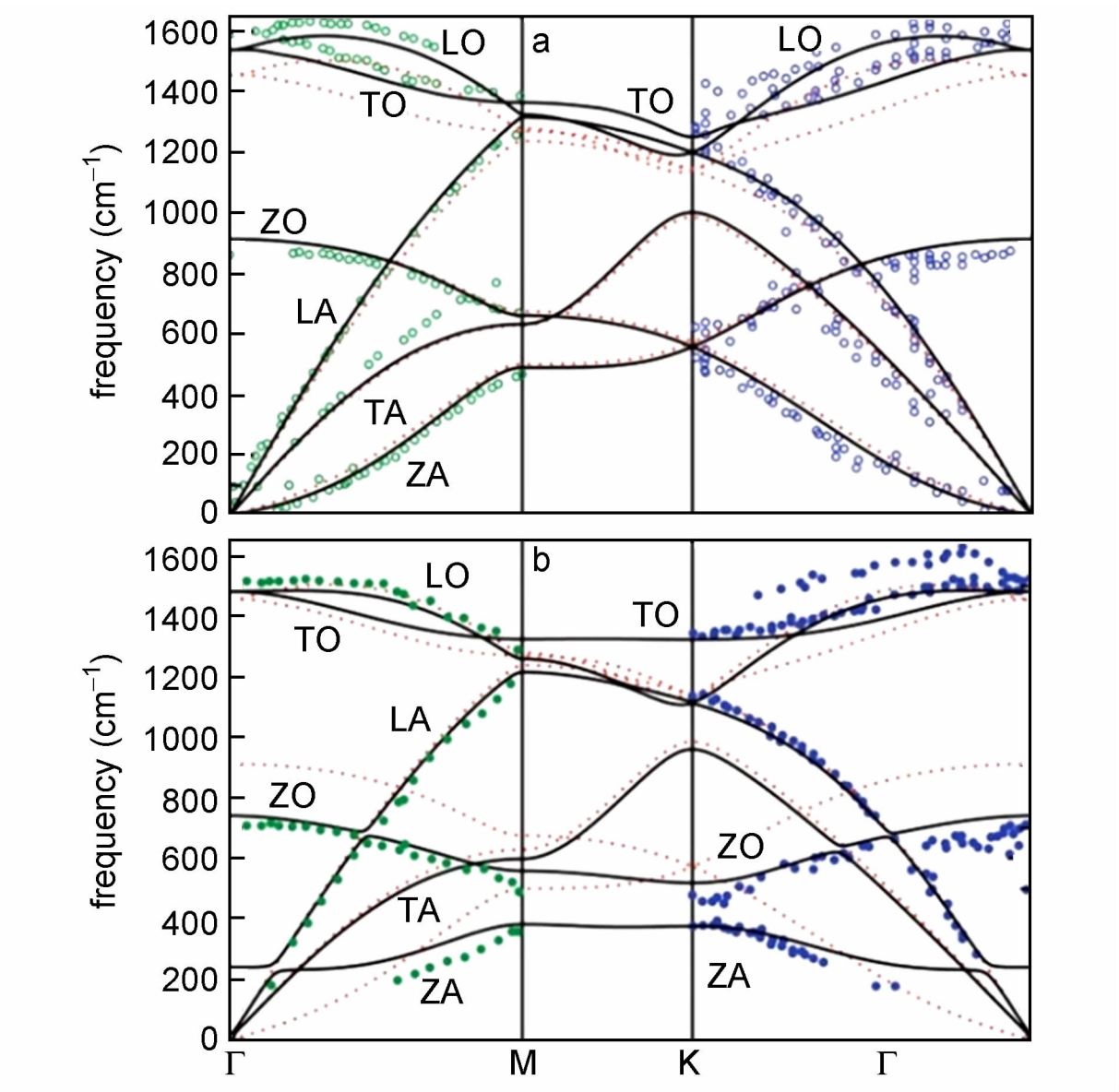

图 10　通过 DFT-LDA 计算的非支撑石墨烯和 Ni 衬底上石墨烯的声子散射[105]

**Fig. 10**　Phonon dispersion of graphene, calculated with DFT-LDA. (a) Isolated graphene: comparison of a calculation using the lattice constant of graphite (black solid lines) with a calculation using the lattice constant of nickel (red dotted lines). (b) Calculated dispersion of graphene on Ni(111) (black solid lines, only the frequencies corresponding to vibrations of C atoms are shown). The red dotted lines present the dispersion of isolated graphene using the lattice constant of Ni as in panel (a)[105]

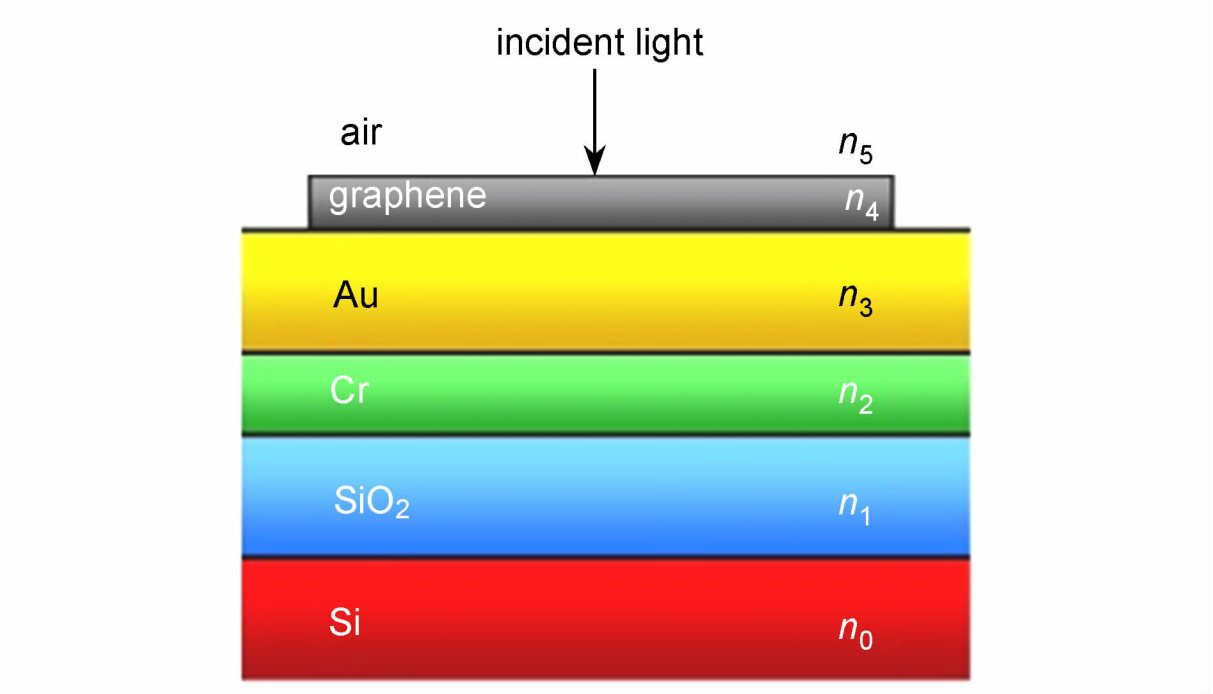

图 11　计算模型中材料在各层的分配[106]

**Fig. 11**　Arrangement of layers used in calculations[106]

## 5　展望

石墨烯具有一系列独特电子性质，对于发展新型纳电子器件有重要意义，是目前国际上的研究热点。大面积、高质量石墨烯制备及其电子结构调控是石墨烯基纳电子器件研究的前提和基础。研究表明，采用化学气相沉积、外延生长等方法可在衬底表面上制备出较大面积、高质量石墨烯材料，并制备出各种石墨烯基原型晶体管、传感器、超级电容器等，揭示出石墨烯-衬底体系的优异性能和潜力。然而对于石墨烯和衬底材料间相互机理、衬底对石墨烯改性的原理及调控、石墨烯-衬底和复合物之间的性能开发等仅有少量报道，对衬底上石墨烯电子结构调控机理缺少系统性认识，这给衬底上功能化石墨烯的进一步研究带来困难。认知衬底对石墨烯电子结构调控的机理，进而设计具有良好电子特性的石墨烯基纳器件结构模型，满足其在纳电子器件中的应用，将是今后研究的热点。

## 参 考 文 献